\RequirePackage{fix-cm}
\documentclass[twocolumn]{svjour3}          % twocolumn
\smartqed  % flush right qed marks, e.g. at end of proof
\usepackage[english]{babel}
\usepackage{graphicx}
\usepackage{makecell}
\usepackage{adjustbox}

\RequirePackage{lineno}
\usepackage{lineno}
\journalname{Computing and Software for Big Sciences}
\begin{document}

\title{Using ATLAS@Home to exploit extra CPU from busy grid sites}

\subtitle{}

\author{Wenjing Wu \and David Cameron \and Di Qing}
\institute{Wenjing Wu
\at Institute of High Energy Physics, CAS, 19B Yuquan Road, Beijing, 100049, China\\Tel.: +8610-88236883, Fax: +8610-88236839\\\email{wuwj@ihep.ac.cn}
\and
David Cameron
\at Department of Physics, University of Oslo, P.b. 1048 Blindern, N-0316 Oslo, Norway
\and
Di Qing
\at TRIUMF, Vancouver, BC, V6T2A3 Canada
}

\date{Received: date / Accepted: date}
\maketitle

\begin{abstract}
Grid computing typically provides most of the data processing resources for large High Energy Physics experiments. However typical grid sites are not fully utilized by regular workloads. In order to increase the CPU utilization of these grid sites, the ATLAS@Home volunteer computing framework can be used as a backfilling mechanism. Results show an extra 15\% to 42\% of CPU cycles can be exploited by backfilling grid sites running regular workloads while the overall CPU utilization can remain over 90\%. Backfilling has no impact on the failure rate of the grid jobs, and the impact on the CPU efficiency of grid jobs varies from 1\% to 11\% depending on the configuration of the site. In addition the throughput of backfill jobs in terms of CPU time per simulated event is the same as for resources dedicated to ATLAS@Home. This approach is sufficiently generic that it can easily be extended to other clusters.

\keywords{BOINC \and ATLAS@Home \and CPU Utilization \and grid site \and backfilling}
%\PACS{PACS code1 \and PACS code2 \and more}
%\subclass{MSC code1 \and MSC code2 \and more}
\end{abstract}

\section{Introduction}
\label{intro}
Large High Energy Physics (HEP) experiments require a huge amount of computing resources for their data processing~\cite{lhc1}\cite{lhc2}. The ATLAS experiment is the largest of the LHC experiments in terms of computing resources and its computing infrastructure~\cite{atlasc1}\cite{atlasc2} is built on grid computing. ATLAS jobs are a mixture of single-core and multi-core~\cite{atlasmcore} workflows which typically use between 4 and 12 cores on a single node (depending on site configuration). The real time computing resources available to ATLAS in 2018 from grid sites are around 2.5 million HEPSPEC06\footnote{HEPSPEC06 is the HEP-wide benchmark for measuring CPU performance and the official CPU performance metric used by the Worldwide LHC Computing Grid. The average performance of one CPU core is around 10 HEPSPEC06 for the ATLAS grid sites.}~\cite{wlcg2018}. ATLAS also uses an increasing level of opportunistic computing resources such as clouds, High Performance Computing\cite{atlasoc} and volunteer computing.

Even though grid sites provide 75\% of the total computing resources to ATLAS, opportunistic computing resources play an important role. One such resource is the volunteer computing project ATLAS@Home\cite{atlasathome1}\cite{atlasathome2} which uses the BOINC\cite{boinc1}\cite{boinc2} middleware to harness worldwide heterogeneous volunteer computers. The ATLAS@Home project is integrated into the ATLAS workload management system PanDA\cite{panda1}\cite{panda2}, and processes ATLAS simulation tasks\cite{atlassimul1}\cite{atlassimul2}. Simulation is a CPU-intensive task which on average consumes over half of the wall time of the ATLAS CPUs.

Most grid sites are clusters managed by batch systems such as HTCondor\cite{htcondor}, SLURM\cite{slurm} and PBS\cite{pbs}, and the scale of the sites ranges from a few hundred to tens of thousands of cores. However, when the CPU time utilization of several ATLAS grid sites was measured, results showed that none of these clusters were being fully used. In other words, both the wall time utilization and CPU time utilization rates were not as high as expected. This means a significant percentage of cluster resources were being wasted, hence the need to seek solutions to improve the CPU time utilization. 

The rest of this paper is organized as follows: Section~\ref{sec:griduse} analyzes the CPU time utilization of the ATLAS grid sites, Section~\ref{sec:usingbackfill} introduces a new method of backfilling the grid sites, Section~\ref{sec:harvest} presents results of backfilling two ATLAS grid sites, Section~\ref{sec:impact} measures the impact of backfilling and Section~\ref{sec:conclusion} concludes.

\section{Utilization of grid sites}
\label{sec:griduse}

\subsection{Analysis from the ATLAS job archive}
In order to understand the utilization rate of grid sites, a few example sites from ATLAS are studied. The selected sites are of different scale and locations and they are dedicated to ATLAS, so the CPU time and wall time of ATLAS jobs is representative of the overall usage of the clusters. CPU efficiency ($\epsilon_{\mathrm{CPU}}$) is used to measure the efficiency of the jobs, and wall time utilization ($u_{\mathrm{wall}}$) and CPU time utilization ($u_{\mathrm{cpu}}$) measure how fully these clusters are being utilized. Assuming that in a given period $M$ days, the total wall time (in seconds) of all jobs is $T_{\mathrm{wall}}$, the total CPU time (in seconds) of all jobs is $T_{\mathrm{CPU}}$, and the total number of available cores of the site is $N_{\mathrm{core}}$, then:

\begin{equation}
u_{\mathrm{wall}} = \frac{T_{\mathrm{wall}}}{3600\times 24 \times M \times N_{\mathrm{core}}}
\end{equation}

\begin{equation}
u_{\mathrm{cpu}} = \frac{T_{\mathrm{cpu}}}{3600\times 24 \times M \times N_{\mathrm{core}}}
\end{equation}

\begin{equation}
\epsilon_{\mathrm{CPU}} = \frac{T_{\mathrm{cpu}}}{T_{\mathrm{wall}}}
\end{equation}

\begin{table}[h!]
\caption{The average utilization of typical ATLAS grid sites over a period of 100 days}
\label{tab:cpugridsites}       
\centering
\begin{adjustbox}{max width=\textwidth}
\begin{tabular}{llllllllll}
%\begin{tabular}{*{5}{|c}|}%%{|c|c|c|c|c|c|c|c|c|c|c|c|c|c|}
\hline\noalign{\smallskip}{\penalty -100 }
Site & \makecell{Amount\\of Cores} &\makecell{ Avg.\\$u_{\mathrm{wall}}$} & \makecell{Avg.\\$u_{\mathrm{wall}}$} &  \makecell{Avg.\\$\epsilon_{\mathrm{CPU}}$} \\
\noalign{\smallskip}\hline\noalign{\smallskip}
BEIJING & 634 &  68\% & 55\% &  81\% \\
TOKYO& 6144&		85\%&	72\%&	85\%\\
SiGNET&	5288&		88\%&	68\%&		77\%\\
MWT2&	16250&		83\%&	70\%&		84\%\\
AGLT2&	10224&		72\%&	61\%&		84\%\\
\noalign{\smallskip}\hline
\end{tabular}
\end{adjustbox}
\end{table} 

As shown in Table~\ref{tab:cpugridsites}, 5 ATLAS sites were chosen from Asia, North America and Europe. They have different scales in terms of the number of cores, and they use different local batch systems. From the selected sites, the average $u_{\mathrm{wall}}$ is around 85\%, and the corresponding $u_{\mathrm{cpu}}$ is around 70\%. Ideally, $u_{\mathrm{wall}}$ should be close to 100\%, but there are several reasons why grid sites cannot achieve this, as follows.

(1)	Sites often have downtime for scheduled maintenance or unexpected problems.

(2)	The inefficiency of both the grid scheduling system and local batch systems. In the ATLAS case, the central PanDA scheduling system is rather conservative, and sites are assigned fewer jobs during the periods before and after downtimes.

(3) Over 50\% of the ATLAS worker nodes run multi-core jobs which have lower CPU efficiency compared to the single-core jobs. This is due to the fact that certain stages of the multi-core job can only use a single core and hence leave the other allocated cores idle.

(4) Sites with fixed partitioning of worker nodes between single-core and multi-core ATLAS jobs can have idle worker nodes when the mix of workloads assigned to the site does not well match the partition well.

(5) For sites configured to mix single and multi-core jobs on the same worker nodes, the multi-core jobs may need to wait for a number of single-core jobs to finish in order to obtain the number of cores they require.

In the best case, even if the site has 100\% $u_{\mathrm{wall}}$, $u_{\mathrm{cpu}}$ would still be less than 100\% because the CPU efficiency of the jobs is always less than 100\%, so the CPU time utilization is always lower than wall time utilization. Different types of job demonstrate different CPU efficiency.

\begin{figure*}[h!]
\includegraphics[width=0.5\textwidth]{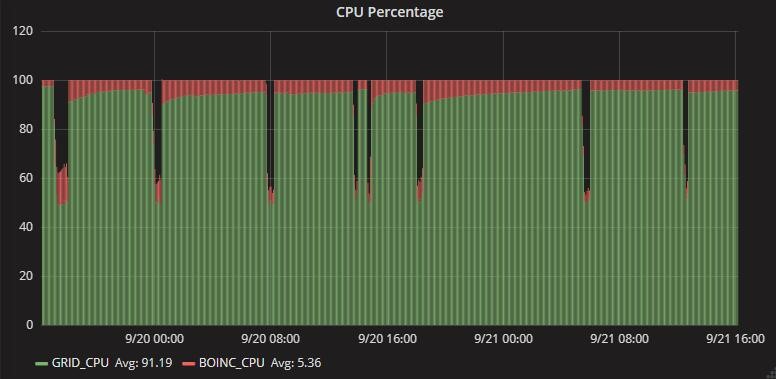}
%\caption{CPU utilization in one day}
\includegraphics[width=0.5\textwidth]{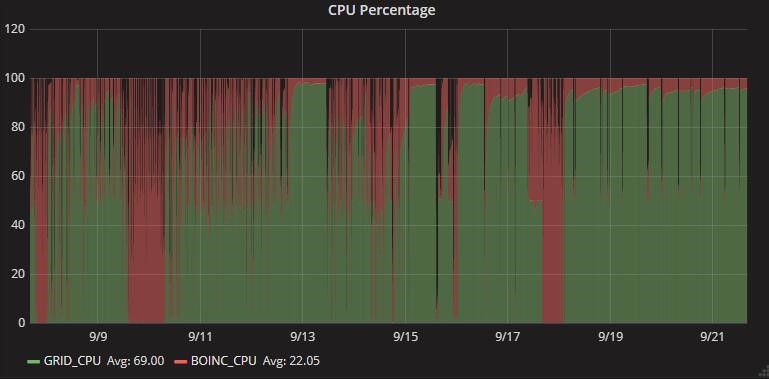}
\caption{CPU utilization on one node over one day (left) and two weeks (right). Green: grid jobs, red: BOINC jobs}
\label{fig:1}       % Give a unique label
\end{figure*}

\subsection{Observation from site's local monitoring}
Using local monitoring tools to look at the CPU time utilization of single worker nodes in different periods, it was observed that in the long run, the CPU time utilization of the worker nodes was not as high as expected.

As shown in Fig.~\ref{fig:1}, on a worker node for the ATLAS BEIJING site, the CPU time utilization of grid jobs (in green) can reach 91\% over a 24 hour period, because this worker node is running highly CPU efficient simulation jobs. But on the same worker node, looking over a period of two weeks, the CPU time utilization is only 69\%. This is because the site had two scheduled downtimes in those two weeks, and also because of the inefficiency of the job scheduling and the jobs. 

\section{Using ATLAS@Home to backfill the sites}
\label{sec:usingbackfill}
\subsection{The basic idea}
From section~\ref{sec:griduse}, it can be seen that with the traditional batch system assignment of one job slot per core, the CPU cycles can never be 100\% utilized due to the job CPU efficiency. The key is to have more than one job slot on each core, but jobs must have different priorities, otherwise more wall time and CPU time would be wasted on the scheduling of CPU cycles between different jobs at the operating system level. In addition, sites use different batch systems so it is not easy to implement a universal configuration for all batch systems, and some batch systems may not support the feature of defining more than one job slot per core and assigning different priorities to different jobs.

Using ATLAS@Home meets the above requirements in terms of being independent from the sites' local batch system and having the ability to use different job priorities.Using the ATLAS@Home platform to run ATLAS@Home jobs in the background of the regular grid job workload effectively exploits CPU cycles which can not be fully utilized by the grid jobs.

\subsection{The advantages of ATLAS@Home jobs}
When ATLAS@Home started it was aimed towards the general public, most of whom were running hosts with the Microsoft Windows operating system. Therefore it was developed to use virtualization to provide the required Linux-based computing environment (operating system, and dependent software installation). Later, as more and more Linux hosts joined the project, containerization and native running were developed to replace virtualization on Linux hosts. This improved the average CPU efficiency of the ATLAS@Home jobs by up to 10\% and is also more lightweight to deploy as it does not require the pre-installation of virtualization software.

Like many volunteer computing projects, the ATLAS@Home project uses the BOINC middleware to manage job distribution to volunteer hosts. A BOINC project defines jobs in a central server, and volunteers install the BOINC client software and configure it to pull jobs from the servers of the projects to which they would like to contribute. A grid site wishing to run ATLAS@Home installs the BOINC client on its worker nodes and configures it to take jobs from the ATLAS BOINC server. In this paper ``BOINC jobs'' are defined as the jobs which BOINC controls on a worker node (as opposed to grid jobs controlled by a batch system), whereas ATLAS@Home is the general framework for volunteer computing in ATLAS.

One key feature of BOINC is that the processes are set to the lowest priority in the operating system, so they only use CPU cycles when they are not being used by any other higher priority processes. In particular, for Linux systems it uses the non-preempt scheduling\cite{nonpe} mechanism for CPU cycles, which means the higher priority processes will always occupy the CPU unless they voluntarily release it. This feature guarantees that starting low priority processes, such as all the processes spawned by the BOINC jobs, will not increase the wall time of the higher priority processes due to switching CPU cycles between processes. Hence BOINC should not impact the CPU efficiency of the higher priority grid jobs. Of course, the CPU efficiency might be lower due to the memory contention of both jobs (overflowing of memory into swap space can prolong the wall time of the jobs).

Another advantage of using BOINC to add the extra job slots is that these jobs are from two different batch systems: the higher priority jobs from the local batch system of the cluster, and the lower priority jobs from BOINC. They are invisible to each other, and the local batch system does not know the BOINC jobs exist, so it will still send as many jobs as it is configured to. In other words, this does not affect the wall time utilization of the higher priority grid jobs.

BOINC provides a convenient way to schedule payloads to the worker node because it is already fully integrated into ATLAS distributed computing systems. Alternative methods of over-committing resources would require either requesting sites to re-configure batch systems to allow over-commit, or developing a way to schedule jobs behind the batch system - essentially duplicating BOINC's functionality.

The multi-core simulation jobs of ATLAS@Home use very little memory (less than 300 MB per core for 12-core jobs), and the majority of ATLAS grid jobs (except for special jobs requiring higher memory) use less than 1.5GB memory per core. This means that grid jobs and BOINC jobs usually have enough memory to co-exist on the same worker node, and the BOINC jobs can also be kept in memory while they are suspended (if for example no CPU cycles are available). Therefore the BOINC jobs do not get preempted even if the grid jobs are using 100\% of the CPU, hence no CPU cycles are wasted.

There is on-going work to integrate ATLAS@Home with the ATLAS Event Service~\cite{aes}, a framework which reduces the granularity of processing from the job-level to the event-level. Events are uploaded to grid storage as they are produced which make it ideal for opportunistic resources where jobs may be terminated at any point. For ATLAS@Home it will be useful in cases where memory requirements are tighter and BOINC jobs cannot be held in memory, so that when a BOINC job is preempted only the current event being processed is lost.

% Suggestion from Rod to remove this section and keep it for future paper on dynamic boinc configuration
%\subsection{Monitoring and statistics}
%In order to keep track and quantify the extra CPU resources being exploited by the BOINC backfilling jobs and measure the impact on the grid jobs, we use the ATLAS Elastic Search and Kibana\cite{EK} analytic platform to create visualizations and perform analysis. Each ATLAS job (including grid jobs and ATLAS@Home jobs) has its information stored in the analytic platform, including the CPU time, wall time, and number of events etc. Also in order to check the details of every single worker node, we deploy a local agent which collects information such as the number of each kind of jobs, the CPU utilization and memory usage for each kind of jobs, system load, swap usage etc. The local information is stored into Elastic Search, and can be visualized via Grafana\cite{grafana}. We also deploy cron jobs to run on the worker nodes, which analyses the output of the local agent, including they system health status (system load, swap usage) and BOINC activity (memory usage by BOINC jobs, number of jobs from BOINC), then take actions accordingly, such as aborting the BOINC jobs if necessary.

\section{The harvest from the grid sites}
\label{sec:harvest}
The ATLAS@Home backfill method was tested on two ATLAS grid sites. The first is a small site in China (BEIJING) which has 464 cores and PBS as its batch system, and the second is a large site in Canada (TRIUMF) which has 4816 cores and HTCondor as its batch system. Both sites are dedicated to ATLAS, so the ATLAS job measurements can serve as an overall measure of the sites' efficiency. The BOINC software was deployed on both clusters, and the worker nodes received jobs from ATLAS@Home to run in the background while the grid jobs were also running. In order to compare the difference, the CPU time utilization and wall time utilization defined in section 2.1 are used.

\subsection{Results from the BEIJING site}
Backfilling was started on the BEIJING site in September 2017. Results from both ATLAS job monitoring and local monitoring during this period suggest that the CPU time exploited by BOINC is dependent on the wall time and CPU time utilization of the grid jobs. In addition to the $u_{\mathrm{cpu}}$, $u_{\mathrm{wall}}$ and $\epsilon_{\mathrm{CPU}}$ metrics defined in section 2.1, an additional metric $f_s$ was used to measure the effect of BOINC jobs on the success rate of grid jobs. $f_s$ is defined as the ratio between successful jobs and total jobs.

Tables~\ref{tab:beijingbusy} and~\ref{tab:beijingidle} show the utilization of BOINC, Grid and All jobs over two different periods of 7 days. In a busy week, the average $u_{\mathrm{wall}}$ of the grid jobs reaches 93\%, and the corresponding $u_{\mathrm{cpu}}$ is 80\%. Under these circumstances, BOINC backfilling jobs can exploit an extra 15\% CPU time from the cluster, which makes the average overall $u_{\mathrm{cpu}}$ of the cluster reach 95\%. With backfilling jobs, the average overall $u_{\mathrm{wall}}$ is 181\%, which means there are on average 1.81 ATLAS processes running or waiting on each core.

In an idle week, the $u_{\mathrm{wall}}$ of the grid jobs is only 62\%, and the corresponding $u_{\mathrm{cpu}}$ of grid jobs is 48\%. In this case, the BOINC backfilling jobs exploit an extra 42\% CPU time, which makes the overall $u_{\mathrm{cpu}}$ of the cluster reach 90\%.

It can be seen that BOINC backfilling can exploit the CPU cycles which cannot be used by grid jobs, and the $u_{\mathrm{cpu}}$ of BOINC jobs depends on the $u_{\mathrm{cpu}}$ of the grid jobs. In addition the overall $u_{\mathrm{cpu}}$ also depends on the $u_{\mathrm{cpu}}$ of the grid jobs, usually higher $u_{\mathrm{cpu}}$ of grid jobs yields higher overall $u_{\mathrm{cpu}}$; For 6 months in BEIJING, the average overall $u_{\mathrm{cpu}}$ of the site remains above 85\%.

\begin{table}
\caption{Utilization of BEIJING site in a busy week}
\label{tab:beijingbusy}       % Give a unique label
\centering
\begin{tabular}{lllll}
\hline\noalign{\smallskip}
 & $f_s$ & $\epsilon_{\mathrm{CPU}}$ & $u_{\mathrm{cpu}}$ & $u_{\mathrm{wall}}$ \\
\noalign{\smallskip}\hline\noalign{\smallskip}
BOINC & 1.00 & 0.17 & 0.15 & 0.88 \\
Grid & 0.99 & 0.53 & 0.80 & 0.93 \\
All & 0.99 & 0.53 & 0.95 & 1.81 \\
\noalign{\smallskip}\hline
\end{tabular}
\end{table}

\begin{table}
\caption{Utilization of BEIJING site in an idle week}
\label{tab:beijingidle}       % Give a unique label
\centering
\begin{tabular}{lllll}
\hline\noalign{\smallskip}
 & $f_s$ & $\epsilon_{\mathrm{CPU}}$ & $u_{\mathrm{cpu}}$ & $u_{\mathrm{wall}}$ \\
\noalign{\smallskip}\hline\noalign{\smallskip}
BOINC & 1.00 & 0.47 & 0.42 & 0.88 \\
Grid & 0.96 & 0.61 & 0.48 & 0.62 \\
All & 0.98 & 0.61 & 0.90 & 1.50 \\
\noalign{\smallskip}\hline
\end{tabular}
\end{table}

\begin{table}
\caption{Utilization of TRIUMF site before backfilling}
\label{tab:triumfbefore}       % Give a unique label
\centering
\begin{tabular}{lllll}
\hline\noalign{\smallskip}
 & $f_s$ & $\epsilon_{\mathrm{CPU}}$ & $u_{\mathrm{cpu}}$ & $u_{\mathrm{wall}}$ \\
\noalign{\smallskip}\hline\noalign{\smallskip}
BOINC & n/a & n/a & n/a & n/a \\
Grid & 0.90 & 0.80 & 0.69 & 0.88 \\
All & 0.90 & 0.80 & 0.69 & 0.88 \\
\noalign{\smallskip}\hline
\end{tabular}
\end{table}

\begin{table}
\caption{Utilization of TRIUMF site after enabling backfilling}
\label{tab:triumfafter}       % Give a unique label
\centering
\begin{tabular}{lllll}
\hline\noalign{\smallskip}
 & $f_s$ & $\epsilon_{\mathrm{CPU}}$ & $u_{\mathrm{cpu}}$ & $u_{\mathrm{wall}}$ \\
\noalign{\smallskip}\hline\noalign{\smallskip}
BOINC & 0.97 & 0.29 & 0.27 & 0.91 \\
Grid & 0.95 & 0.50 & 0.65 & 0.97 \\
All & 0.95 & 0.50 & 0.92 & 1.88 \\
\noalign{\smallskip}\hline
\end{tabular}
\end{table}

\subsection{Results from the TRIUMF site}
For the TRIUMF site, the overall $u_{\mathrm{cpu}}$ of the site before and after adding the BOINC backfilling jobs is compared.

Table~\ref{tab:triumfbefore} shows a 7-day period before adding the backfilling jobs, during which the average overall $u_{\mathrm{cpu}}$ is 69\%. Table~\ref{tab:triumfafter} shows a 7-day period when backfilling was enabled, when the average overall $u_{\mathrm{cpu}}$ is 92\% of which 27\% is exploited by the backfilling jobs. It is also notable that the average $u_{\mathrm{wall}}$ of grid jobs after is 9\% higher, in other words the backfilling jobs do not affect the throughput of the grid jobs; After adding the backfilling jobs the overall $u_{\mathrm{wall}}$ of the cluster is 188\%, corresponding to an average 1.88 ATLAS processes running or waiting on each core.  

\section{Measuring the effects of backfilling}
\label{sec:impact}
In order to understand the impact of the backfilling jobs on the grid jobs and vice-versa, several metrics are used to compare them: the $\epsilon_{\mathrm{CPU}}$ and $f_s$ defined respectively in section 2.1 and 4 for grid jobs, and the CPU time per event for the BOINC jobs.

\subsection{Failure of grid jobs}
Tables~\ref{tab:beijingbusy}-\ref{tab:triumfafter} show that the $f_s$ of jobs for both sites remains very high after adding the backfilling jobs. In fact, the $f_s$ is even 5\% higher for TRIUMF after adding the backfilling jobs, indicating that the backfilling jobs do not have any negative effect on the grid job success rate. 

\subsection{CPU efficiency of grid jobs}
To study the effect of backfilling on CPU efficiency of grid jobs, a reliable and stable set of jobs needed to be found. Rather than using all the ATLAS jobs over a certain period of time, only simulation jobs whose wall time was longer than 0.3 CPU days were selected. There were several reasons for this: simulation jobs on average use over 50\% of a site’s CPU time, there is usually a constant flow of them over time, and these jobs have much higher and more stable $\epsilon_{\mathrm{CPU}}$ compared to the other types of ATLAS jobs. In addition, restricting to jobs longer than 0.3 CPU days leads to average $\epsilon_{\mathrm{CPU}}$ above 95\% and increases the sensitivity of the measurement of the effect of backfilling.

Table~\ref{tab:2} shows the average $\epsilon_{\mathrm{CPU}}$ for 6 sets of simulation tasks (3 before running backfill, 3 after) running on the BEIJING site. The jobs all used 12 cores. The $\epsilon_{\mathrm{CPU}}$ of grid simulation jobs drops by between 1.12\% and 1.92\% after adding the backfilling jobs. This is expected, as a little bit of extra wall time can be added to the grid jobs if there is memory contention between the grid and BOINC jobs.  

When comparing the $\epsilon_{\mathrm{CPU}}$ in TRIUMF, the difference is larger. As shown in Table~\ref{tab:3}, the $\epsilon_{\mathrm{CPU}}$ of grid simulation jobs drops by between 10.02\% and 13.32\% after adding the backfilling jobs. The drop can mainly be ascribed to two reasons. Firstly the memory usage of grid jobs in TRIUMF is higher since it runs 6-core multi-core jobs compared to 12-core in BEIJING. TRIUMF also runs a larger variety of ATLAS jobs, some of which have higher memory requirements. Secondly, TRIUMF uses cgroups~\cite{cgroup} to control the resource allocation between grid and BOINC jobs. With cgroups, BOINC jobs could ``steal'' the CPU cycles from the grid jobs, in other words, with cgroups BOINC is allocated more CPU cycles than it should have been.
 
\begin{table}[h!]
\caption{CPU efficiency comparison for grid jobs in BEIJING site (12 cores per job)}
\label{tab:2}       
%\centering
\begin{adjustbox}{max width=0.5\textwidth}
\begin{tabular}{llllllllll}
%\begin{tabular}{*{6}{|c}|}%%{|c|c|c|c|c|c|c|c|c|c|c|c|c|}
\hline\noalign{\smallskip}{\penalty -100 }
 & \makecell{Sample\\jobs} &\makecell{ Avg. MEM\\(MB)per core} & \makecell{Avg. $\epsilon_{\mathrm{CPU}}$\\(\%) per core} &  \makecell{Avg. wall time\\(day)} \\
\noalign{\smallskip}\hline\noalign{\smallskip}
Before  & 113 & 405.04 & 97.07 & 0.44 \\
Before & 387& 402.77 & 97.23 & 0.58 \\
Before &	430& 403.44 & 97.37 & 0.52 \\
After & 127&	394.95 & 95.95 & 0.64 \\
After &	292& 374.24 & 95.88 & 0.68 \\
After &	120& 389.12 & 95.45 & 0.41 \\
\noalign{\smallskip}\hline
\end{tabular}
\end{adjustbox}
\end{table} 

\begin{table}[h!]
\caption{CPU efficiency comparison for grid jobs in TRIUMF site (6 cores per job)}
\label{tab:3}       
%\centering
\begin{adjustbox}{max width=0.5\textwidth}
\begin{tabular}{llllllllll}
%\begin{tabular}{*{6}{|c}|}%%{|c|c|c|c|c|c|c|c|c|c|c|c|c|}
\hline\noalign{\smallskip}{\penalty -100 }
 & \makecell{Sample\\jobs} &\makecell{ Avg. MEM\\(MB)per core} & \makecell{Avg. $\epsilon_{\mathrm{CPU}}$\\(\%) per core} &  \makecell{Avg. wall time\\(day)} \\
\noalign{\smallskip}\hline\noalign{\smallskip}
Before  & 79 & 248.38 & 97.67 & 0.60 \\
Before & 259& 550.98 & 97.61 & 0.62 \\
Before & 2534& 541.40 & 97.59 & 0.41 \\
After & 542&	542.21 & 87.65 & 0.61 \\
After &	168& 541.78 & 84.35 & 0.69 \\
After &	2858& 539.72 & 86.36 & 0.59 \\
\noalign{\smallskip}\hline
\end{tabular}
\end{adjustbox}
\end{table} 

However, this is tunable from both the BOINC and site's resource allocation, depending on whether the goal of the site is to maximize the overall CPU time utilization of the cluster or to minimize the $\epsilon_{\mathrm{CPU}}$ drop of the grid jobs. In general, since both grid and backfilling jobs are ATLAS jobs, for ATLAS dedicated sites, it is obvious that the goal should be to maximize the overall CPU time utilization.

\begin{table*}[ht!]
\caption{CPU time per event comparison for BOINC jobs}
\label{tab:4}       
\centering
%\begin{adjustbox}{max width=0.5\textwidth}
\begin{tabular}{llllllllll}
%\begin{tabular}{*{8}{|c}|}%%{|c|c|c|c|c|c|c|c|c|c|c|c|c|c|}
\hline\noalign{\smallskip}{\penalty -100 }
 Task& \makecell{Dedicated\\Sample\\jobs} & \makecell{Dedicated\\cpu(sec)\\per event} &\makecell{Dedicated\\$\epsilon_{\mathrm{CPU}}$(\%)} &  \makecell{Backfilling\\Sample\\jobs} & \makecell{Backfilling\\cpu(sec)\\per event} &\makecell{Backfilling\\$\epsilon_{\mathrm{CPU}}$(\%)} & \makecell{offset(\%)\\CPU time\\per event} \\
\noalign{\smallskip}\hline\noalign{\smallskip}
1& 673 & 172.02 & 91.49 & 3235 & 165.59 & 34.66 & 4\\
2&	15&	225.21&	93.37&	241&	219.68&	31.69&	2\\
3&	59&	255.41&	93.96&	320&	246.82&	48.76&	3\\
4&	255&	200.99&	91.90&	1220&	198.55&	34.30&	1\\
5&	74&	211.48&	92.98&	334&	204.66&	38.26&	3\\
6&	60&	289.73&	93.85&	320&	291.78&	43.38&	1\\
7&	78&	481.49&	95.06&	284&	471.89&	48.53&	2\\
8&	248& 218.78&	93.01&	596&	220.32&	51.00&	1\\
\noalign{\smallskip}\hline
\end{tabular}
%\end{adjustbox}
\end{table*} 

\subsection{Impact of backfilling on ATLAS@Home}
The effects on running BOINC jobs in backfill mode can be measured by comparing similar jobs running on dedicated (BOINC-only) nodes and backfill nodes which have the same hardware configuration. The following results came from one set of 48 cores dedicated for BOINC jobs and another set of 400 cores which ran both grid jobs and backfilling jobs. The metric used for comparison is the consumed CPU time per simulation event processed (a BOINC job consists of processing 200 events). 

Since jobs from the same simulation task take a similar time to simulate each event, 8012 sample jobs from 8 different simulation tasks were selected to compare the dedicated and backfill nodes. As shown in Table~\ref{tab:4}, for each task the CPU time per event for the BOINC jobs differs by only 1-4\% between the dedicated and backfill cores. This indicates that the CPU time exploited by the BOINC backfilling jobs (when they are actually using CPU) is similar to the CPU time from dedicated nodes. The $\epsilon_{\mathrm{CPU}}$ is a clear indicator of whether the job is run on dedicated or backfilling cores - $\epsilon_{\mathrm{CPU}}$ for backfilling jobs is much lower because they have to wait for CPU cycles to be released by higher priority processes.

\section{Conclusion}
\label{sec:conclusion}
There are many factors causing low overall CPU efficiency of grid sites, and this study shows that for ATLAS grid sites it is very difficult to achieve CPU time utilization above 70\% of the CPU time available from the site. The ATLAS@Home framework provided a convenient solution to experiment with backfilling grid sites thanks to a few unique and convenient features of the ATLAS@Home jobs. Running BOINC backfilling jobs on two ATLAS grid sites (one small site and one medium size site) has demonstrated that using backfilling can exploit a considerable amount of extra CPU time which could not otherwise be used by grid jobs. With backfilling jobs, the overall CPU time utilization reaches over 90\% for both sites. This improves the overall CPU time utilization of the cluster by 15-42\% depending on the workload of the grid jobs. The impact of the backfilling jobs was also measured. From the grid jobs’ point of view, there is no impact on the failure rate. The impact on the CPU efficiency of grid jobs is 1-11\% depending on the configuration of the site, the memory usage of grid jobs and the resource allocation configuration. From the BOINC jobs’ point of view, the CPU time exploited in the backfilling model generates the same amount of events as the CPU time from resources dedicated to BOINC.

Based on both the improvement of the overall CPU time utilization of the site and the impact on the CPU efficiency on the grid jobs, for the sites dedicated to ATLAS it is recommended to prioritize the improvement of the overall CPU time utilization over the sacrificing of CPU efficiency of grid jobs. For non-dedicated sites, the BOINC resource allocation can be tuned to balance the overall CPU time utilization improvement and the sacrificing of the CPU efficiency of higher priority jobs. This method has so far been deployed on ATLAS grid sites, but the approach and results could also be extended to general purpose clusters.

\begin{acknowledgements}
This work was done as part of the distributed computing research and development programme within the ATLAS Collaboration, which we thank for their support. In particular we wish to acknowledge the contribution of the ATLAS Distributed Computing team (ADC). This project is supported by the Chinese NSF grants "Research on fine grained Event Service for the BESIII offline software and its scheduling mechanism (No.11675201)" and "Research on BESIII offline software and scheduling mechanism on desktop grid （No.11405195）". We would also like to thank all the volunteers of ATLAS@Home who made this project possible, and also for the support of NCRC, CFI and BCKDF (Canada) for the TRIUMF Tier1 site. ATLAS@Home relies on many products that comprise the ATLAS distributed computing ecosystem and so we would like to acknowledge the help and support of PanDA, Rucio and NorduGrid ARC.
\end{acknowledgements} 



\end{document}